\documentclass[pra,rsi,amsmath,amssymb,reprint]{revtex4-1}

\usepackage{graphicx}
\usepackage{dcolumn}
\usepackage{bm}
\usepackage{natbib}
\setcitestyle{square}

\usepackage{amsmath}
\usepackage{mathrsfs}
\usepackage{amsfonts}
\usepackage{color}
\usepackage[normalem]{ulem} 

\usepackage{hyperref}
\usepackage{cleveref}

\begin{document}

\preprint{APS/123-QED}

\title{Single-particle Relaxation Time in Doped Semiconductors beyond the Born Approximation}

\author{Gionni Marchetti}
\email{gionni.marchetti@kbfi.ee}
\affiliation{%
National Institute of Chemical Physics and Biophysics, R{\"a}vala 10,  10143, Tallinn, Estonia\\
}%




\date{\today}

\begin{abstract}
We compare the  magnitudes of  the single-particle relaxation time  exactly computed  in the  variable phase approach  with those computed in the first Born approximation for  doped semiconductors such as Si and GaAs, assuming that the Coulomb impurities are randomly distributed centers. We find that for typical dopant concentrations in Si,  the Born approximation can overestimate the single-particle relaxation time by  roughly  40\% and underestimate it by  roughly  30\%. Finally, we show that strong interference of phase shifts is missing in strong scattering regime where the Born approximation fails.
\end{abstract}

\maketitle

\section{Introduction} \label{intro}

The first Born approximation (B1) \cite{taylor1972}, hereinafter simply referred to as the Born approximation, is commonly used within condensed matter theory for computing scattering-related properties, e.g. the carrier mobility or the momentum relaxation time. In combination with the random phase approximation (RPA)\cite{giuliani2005}, the B1 provides the theoretical framework for understanding most of quantum processes which occur in bulk semiconductors \cite{ridley2013}. While nothing can be stated \emph{a  priori} about the validity of the  Born approximation for the low energy processes,  it can be certainly assumed good enough for high-energy collisions. Thus, for scattering processes in solid-state systems, in principle  one should calculate several terms of the Born series  to make sure it converges, thereby showing that its first term (which defines the B1) is sufficiently accurate for the problem at hand. This difficult task can be accomplished by the traditional techniques of scattering phase shifts. Unfortunately such calculations can be very tedious and prone to errors  \cite{schiff1968}. 
Several authors have used these traditional quantum mechanical approaches to accurately compute transport properties such electron (or hole) mobilities in  in bulk semiconductor materials, see for instance Refs. \cite{krieger1968, meyer1981, lowney1991, bennett1992}. In this work, we  show that  we can easily depart from B1 for the case of the single-particle relaxation time  by means of the variable phase  method (VPM) \cite{calogero1967}. This is an alternative approach to the computation of the phase shifts, whereby one casts the  Schr\"{o}dinger equation for the scattering problem into a first order nonlinear equation. This approach allows for fast and accurate computations of the phase shifts at low computational cost.
 
Our computations of the single-particle relaxation times for {\it n}-type Si and GaAs semiconductors  improve on those based on the Born approximation for the doping densities under scrutiny, showing that the inaccuracies can be of about 30\% - 40\% for certain dopant concentrations in Si, while they prove to be less severe for GaAs, where they are smaller than about 20\%. In doing so, we also gain some important physical insights about the validity of the screened electron-impurity interaction as typically accounted for at RPA level. In fact, we find that the large inaccuracies arise where  the random phase approximation starts to break down. Moreover we show that in the strong Thomas-Fermi screening limit in Si, due to the failure of B1, the ratio between the scattering time and 
the single-particle relaxation time is much less than unity while it is  expected to be unity  \cite{dasSarma1985}. We argue that the Born approximation in this regime neglects the interference between the partial waves altogether. 
This paper is organized as follows. In Section~ \ref{sec:relaxation},  we discuss  the relation between the  single-particle relaxation time and the Born approximation. In  Section \ref{sec:method}, we present the  variable phase  method in a pedagogical way as it seems  not to be very well-known among  condensed matter theorists. In Section \ref{sec:results}, we present our findings for the single-particle relaxation time for {\it n}-type Si and {\it n}-type GaAs bulk semiconductors within the homogeneous electron gas paradigm (jellium model) and Thomas-Fermi screening. Finally, in Section \ref{sec:results1}, we show that in Si the Born approximation neglects the important interference effect of the partial waves in a doping region of strong screening where this approximation fails.

\section{The Single-particle Relaxation Time for the Thomas-Fermi Screening  in the Born Approximation} \label{sec:relaxation}

In normal metals and degenerate doped semiconductors, the scattering by  static impurities is characterized by two different (electronic) momentum relaxation times  \cite{dasSarma1985}: the scattering time  $\tau_t$  and the single-particle relaxation time  $\tau_s$ . The latter is also commonly known as the quantum lifetime, and in semiconductor physics it is denoted  by $\tau_{nk}$, where $n$ and $\mathbf{k}$ are the band index and the Bloch electron wave vector in the Brillouin zone (BZ) respectively. This quantity is inversely proportional to the imaginary part of the self-energy Im$\Sigma$, i.e., $1/\tau_s=(2/\hbar)$Im$\Sigma$ \cite{dasSarma1985}.

Assuming that in the bulk of solids the impurities are Coulomb  centers randomly distributed,  $\tau_s$  can be computed by the perturbation theoretic approach using the one-electron Green's function for the coupled electron-impurity system \cite{doniach1998}. By this approach, averaging over the spatial distribution of the impurity centers, the single-particle relaxation time  $\tau_s^{1} $ is commonly obtained in B1, and reads \cite{abrikosov1965, dasSarma1985}\footnote{From now on all the quantities with the superscript $1$ are computed in B1, i.e., the superscript refers to  to the first term of the Born series.}  

\begin{equation}\label{eq:tau}
 \frac{1}{\tau_s^{1} } = \frac{2 \pi m^{\ast} n_i}{\hbar^{3}}  \int \frac{d^{3}k'}{\left(2 \pi\right)^{3}}
 | V_{ei}\left(q \right)|^{2}\frac{\delta\left(k'-k_{\mathrm{F}} \right)}{k'}\, ,
\end{equation}
where $n_i$ and $m^{\ast}$  are the dopant concentration  and the electron effective mass, respectively.  The Fourier component of the electron-impurity interaction potential $V_{ei}\left(q \right)$ is computed at the wave vector transfer $q = 2 k_{\mathrm{F}} \sin\left(\theta/2\right)$ where $k_{\mathrm{F}}$
 and $\theta$ are the Fermi wave vector and the scattering angle, respectively \footnote{Assuming that the incident wave and the scattered wave have wave vectors $\mathbf{k}$ and $\mathbf{k'}$ respectively, then $\mathbf{q} = \mathbf{k} -\mathbf{k'}$. Note that $q \equiv |\mathbf{q}|$. }. Hence, according to this physical transport model, the carriers with the Fermi energy $E_\mathrm{F} $ scatter off single-ion sites elastically. We note that the presence of  matrix element squared $ | V_{ei}\left(q \right)|^{2}$ clearly implies that the  scattering amplitude is computed  in B1. Next, in order to prove the last statement,  we  derive the relation between $\tau_s^{1} $  and the total cross-section   $\sigma_{ei, tot}^{1}\equiv \sigma_{ei, tot}^{1}\left(k_{\mathrm{F}} \right) $  evaluated in Born approximation. First, we note that the matrix element $V_{ei}\left(q \right)= \langle \mathbf{k'}_{\mathrm{F}} |V_{ei}|\mathbf{k}_{\mathrm{F}} \rangle$, where $\mathbf{k}_{\mathrm{F}} $ and $\mathbf{k'}_{\mathrm{F}} $ denote the wave vectors, near the Fermi surface, of the incident and scattered plane waves,   respectively,   is equivalent to the Fourier $q$-component of the potential given by \cite{ashcroft1976}

\begin{equation}\label{eq:fourier}
 V_{ei}\left(q \right) = \int d^{3}r e^{i \mathbf{q}\cdot \mathbf{r}}
  V_{ei}\left(r \right)\, .
\end{equation}
Now, the scattering amplitude $f^{1}\left(q \right)$ in B1 can be obtained through   Eq.~\ref{eq:fourier}  and reads \cite{schiff1968}

\begin{equation}\label{eq:amplitude}
 f^{1}\left(q \right) = - \frac{ m^{\ast}}{2 \pi \hbar^2}V_{ei}\left(q \right) \, .
\end{equation}
Note that   Eq.~\ref{eq:amplitude} is meaningful only when the exact scattered wave function is replaced by the incident plane wave. This happens whenever  $V_{ei}$ is sufficiently weak such that  the incident electron plane wave is slightly distorted by its presence. This is the basic idea on which the Born approximation rests. 
Then, the differential cross-section  $\sigma_{ei}^{1}\left(\theta \right)\equiv \sigma_{ei}^{1}\left(k_{\mathrm{F}}; \theta \right)$  in Born approximation can be computed by means of  the relation \cite{taylor1972}
\begin{equation}\label{eq:amplitude1}
  \sigma_{ei}^{ 1}\left(\theta \right)=  |f^{1}\left( 2 k_{\mathrm{F}} \sin\left(\theta/2\right)\right) |^{2} \, .
\end{equation}

Writing  Eq.~\ref{eq:tau} in spherical coordinates, one obtains

\begin{equation}\label{eq:tau1}
 \frac{1}{\tau_s^{1} } = \frac{2 \pi m^{\ast} n_i k_{\mathrm{F}} }{\left(2 \pi \hbar\right)^{3}} \int_{0}^{2\pi }  d\phi  \int_{0}^{\pi }  \sin\theta  \,  d\theta   \, 
 | V_{ei}\left(q \right)|^{2}  \, .
\end{equation}

Combining the double integral in angular coordinates $\theta, \phi$ of Eq.~\ref{eq:tau1} and Eq.~\ref{eq:amplitude1}  one obtains the definition of the total scattering cross-section  $\sigma_{ei, tot}^{1}$ in the Born approximation for the potential  $V_{ei}$. Indeed, the latter scattering quantity reads \cite{taylor1972}
\begin{equation}\label{eq:totalB1def}
  \sigma_{ei,tot}^{1} =  \int_{0}^{2\pi } d\phi  \int_{0}^{\pi }   \sin\theta  \,  d\theta  \,   \sigma_{ei}^{1}\left(\theta \right)  \, .
\end{equation}

From the above formulas, a simple expression for 
the single-particle relaxation time in B1 is readily obtained as \footnote{Note that a very similar formula to Eq.~\ref{eq:tau3}  is obtained in Ref. \cite{abrikosov1965}, but in different units. See  Eq.~  (39.6) at page $326$.}

\begin{equation}\label{eq:tau3}
 \frac{1}{\tau_s^{1} } = \frac{ n_i \hbar k_{\mathrm{F}}   }{m^{\ast}  } \sigma_{ei,tot}^{1} = n_i  v_{\mathrm{F}}   \sigma_{ei,tot}^{1}  \, ,
\end{equation}
where $ v_{\mathrm{F}} = \hbar k_{\mathrm{F}}/m^{\ast} $ is the Fermi velocity. Eq.~\ref{eq:tau3} shows that $\tau_s^{1}$ is inversely proportional to the total cross-section $\sigma_{ei,tot}^{1}$ computed in the Born approximation. Thus, ignoring   the distribution  of the impurity centers in the solid, one can compute the single-particle relaxation time by means  of the total cross-sections which clearly assumes two-body scattering. Therefore, the present formalism breaks down  whenever the electrons scatter off more than one single-ion site at time. In such a case, within the present formalism one may include   some   \emph{ad hoc} empirical corrections such as Ridley's third-body rejection \cite{ridley1977} or others based on more sophisticated models, see Ref. ~ \cite{fischetti2016}. 

So far, we have not yet addressed how  to accurately model the interaction potential between the carriers and the  ionized impurity centers. To this end,  we shall consider a given impurity ion as a test charge embedded into a weakly interacting gas in a paramagnetic state. The linear response theory (LRT) is then applicable, and provides an effective electron-impurity potential  at RPA level \cite{giuliani2005}. We shall limit ourselves to the random phase approximation as for the material parameters under scrutiny exchange and correlation effects are in general negligible as the kinetic energy dominates the exchange and correlation terms. Finally, the last approximation we shall consider  within the LRT,  is that of Thomas-Fermi screening \cite{ashcroft1976, giuliani2005} which gives rise to a RPA effective potential, expected satisfactory for a small momentum transfer, i.e.,  in the limit $q \to 0 $. The latter assumption implies that the form taken by the screened Coulomb potential at RPA, is of Yukawa type \cite{ashcroft1976} (or equivalently of Thomas-Fermi type) and reads as
\begin{equation}\label{eq:fourier1}
V_{ei}\left(q \right) = \frac{4\pi U_0}{q^{2} + q_0^{2}}  \, ,
\end{equation}
which in the direct space is equivalent to the Yukawa potential $U_0 e^{-q_0 r}/r$ of strength $U_0 = - Ze^2/\kappa $ (cgs units), where $e$ is the elementary charge, $Z$ is the impurity valence,  and  $\kappa$ is the background dielectric constant. The Thomas-Fermi wave vector (or inverse screening length) $q_{\mathrm{TF}}$
at $T = 0$ $\mathrm{K}$ (degenerate electron gas) is given by $q _{\mathrm{TF}}^{2}= 6 \pi ne^{2}/\kappa E_\mathrm{F} $ where $n$ is the electron density \cite{giuliani2005}.

For the sake of thoroughness, we recall that the formula for $ \sigma_{ei,tot}^{1} $ when  $V_{ei}$ is given by Eq.~ \ref{eq:fourier1}, reads \cite{schiff1968}

\begin{equation}\label{eq:totalB1}
\sigma_{ei,tot}^{1}= \frac{16 \pi m^{\ast^{2}} U_0^{2}}{\hbar^{4} q _{\mathrm{TF}}^{4} \left(4y^{2} + 1\right)}  \, ,
\end{equation}
where  $y= k_{\mathrm{F}}/q _{\mathrm{TF}}$.  Eq.~\ref{eq:totalB1} illustrates another well-known issue \cite{fischetti1995} of the Born approximation, which  does not distinguish between attractive and repulsive Coulomb interactions. 

From  Eq.~\ref{eq:tau} it becomes evident how to depart from B1 for the single-particle relaxation case. Indeed, one needs to replace $ \sigma_{ei,tot}^{1} $ by the exact total cross-cross section $ \sigma_{ei,tot}$. The latter can be accurately computed by the phase shift formalism, and reads as \cite{taylor1972}
\begin{equation}\label{eq:total1}
\sigma_{ei,tot}= \frac{4 \pi}{k_{\mathrm{F}}^{2}} \sum_{l=0}^{\infty} \left(2l +1 \right) \sin^{2} \delta_l \, ,
\end{equation}
where $\delta_l$ denote the  phase shifts of given angular momentum number $l$, computed for the input Fermi wave number $ k_{\mathrm{F}}$, i.e., $\delta_l \equiv \delta_l\left( k_{\mathrm{F}}\right) $   accordingly to the screened Coulomb potential of Eq.~\ref{eq:fourier1}. 

The phase shifts $\delta_l^{1}$ ($\delta_l^{1}\ll 1$) in B1 are computed by means of \cite{schiff1968}
\begin{equation}\label{eq:delta1}
\delta_l^{1}= -\frac{2 m^{\ast} k_{\mathrm{F}}}{\hbar^{2}} \int_{0}^{\infty } dr [j_l\left(k_{\mathrm{F}} r \right)]^{2} r^{2}  V_{ei}\left(r \right) \, ,
\end{equation}
where $j_l$ are the spherical Bessel functions. Inserting $\delta_l^{1}$  into Eq.~\ref{eq:totalB1}   one would obtain 
the same total cross-section in B1 given in Eq.~\ref{eq:totalB1}. Moreover, for the validity of the Born approximation in term of phase shift, it is expected that $\delta_l^{1} \approx \delta_l$. 

In the following Section, we  shall  perform   the task of computing  $\delta_l$, and hence the exact $\tau_s$,  by means of the  variable phase method \cite{calogero1963,calogero1967,taylor1972}. This approach we would allow us to  accurately compute  the  single particle-relaxation time in doped semiconductors with at relatively small computational cost.

\section{The  Variable Phase  Method} \label{sec:method}

Due to the cylindrical symmetry of the scattering problem, one can expand the electron wave function $\psi \left(r \right) $ by the functions $u_l\left(r \right)$ which are the solutions of the radial Schr\"{o}dinger equation. The latter, setting $2 m^{\ast}$ and to $\hbar$  unity, reads 
\begin{equation}\label{eq:schr}
u''_l  \left(r \right) + \left[k^{2} -l\left(l + 1 \right)/r^{2} - V_{ei}\left(r \right) \right]u_l \left(r \right) = 0 \, .
\end{equation}
The scattering potential $ V_{ei}$ is responsible for the presence of $\delta_l$  in the asymptotic behavior of $u_l\left(r \right)$, i.e. $u_l\left(r \right) \sim \sin  \left(kr - l \pi/2 - \delta_l \right)$  for $ r \to \infty $. 
Thus, solving the radial  Schr\"{o}dinger equation for a scattering problem is equivalent to compute the $\delta_l$, whereby one can obtain the scattering total cross-section, and hence the exact single-particle relaxation time.

The variable phase approach is an alternative method to the  integration of Eq. ~\ref{eq:schr} which directly yields the exact phase shifts except for very small numerical errors.  In VPM one obtains $\delta_l$ by integrating the phase equation, a first order nonlinear equation which is a generalized Riccati equation. The phase equation reads  \cite{calogero1967}
\begin{equation}\label{eq:phase}
\delta'_l \left(r \right) = -  k^{-1} V_{ei}\left(r \right)  \left[\cos\delta_l\left(r \right)	\hat{j_l} \left(kr \right)  - \sin\delta_l\left(r \right)\hat{n}_l \left(kr \right)  \right]^{2}\, , 
\end{equation}
where 	$\hat{j_l}$, $\hat{n}_l$ are the Riccati-Bessel functions and 
 the boundary condition at the origin is given by $\delta_l\left(0\right)=0$ \cite{calogero1967}. The $\delta_l$ numerical values are then defined by the limit
\begin{equation}\label{limit}
\lim_{r \to \infty} \delta_l \left(r \right) = \delta_l \, ,
\end{equation}
where $r$ is the inter-particle distance, that is, taking the asymptotic $\delta_l$  values  far away from the impurity center. 

In order to compute the $\delta_l$  by Eq. ~\ref{eq:phase}, one needs to provide a suitable electron-impurity interaction potential $V_{ei}$ to the radial Schr\"{o}dinger equation. This scattering potential must be short-range and regular enough as typically required in scattering theory, see Refs. ~\cite{calogero1967}\cite{taylor1972}. 

It is fortunate that the Yukawa interaction potential belongs to a class of regular potentials for which one can accurately compute  the true $\delta_l$ by VPM. Throughout this work we shall assume that this is the case along with Ref. ~\cite{caruso2016}, referring the reader to  Refs.~\cite{ashcroft1976,giuliani2005,marchetti2018} for a detailed discussion of its validity within LRT.  
In the following along with Ref. ~ \cite{caruso2016} we shall consider {\it n}-type  Si with  $n_i= 10^{17} - 2.5 \times 10^{20}$ $\mathrm{cm}^{-3} $ taking the longitudinal effective mass  $m_{\parallel }^{\ast} =0.89 m_e $  and the transverse effective mass  $m_{\bot }^{\ast} =0.19 m_e  $ \cite{caruso2016}  \footnote{The overall effective mass is $m^{\ast} =\frac{3}{2/m_{\bot }^{\ast} + 1/m_{\parallel }^{\ast} }  $. i.e. $m^{\ast} \approx 0.25 m_e $. Note that in Ref.~ \cite{madelung2004} $m_{\parallel }^{\ast} =0.91m_e $, however this would make a negligible difference. }, $m_e$  being the bare electron mass,   and $\kappa=12.0\varepsilon_{0}$ is the background dieletric constant  where $\varepsilon_{0}$ is  the vacuum permittivity. Note that for the dopant concentrations under scrutiny, it is usually assumed that the electrons undergo independent scattering processes  with the donors which in turn act as single scattering centers \cite{bennett1992}.

Moreover, we shall consider single-ionized impurities ($Z=1$) and the electron density  will be set to $n = n_i$, and at the same time we shall ignore the  conduction-band valley degeneracy \cite{mahan1980} according to Ref.~\cite{caruso2016}.

\begin{figure}[htp]
\resizebox{0.50\textwidth}{!}{%
  \includegraphics{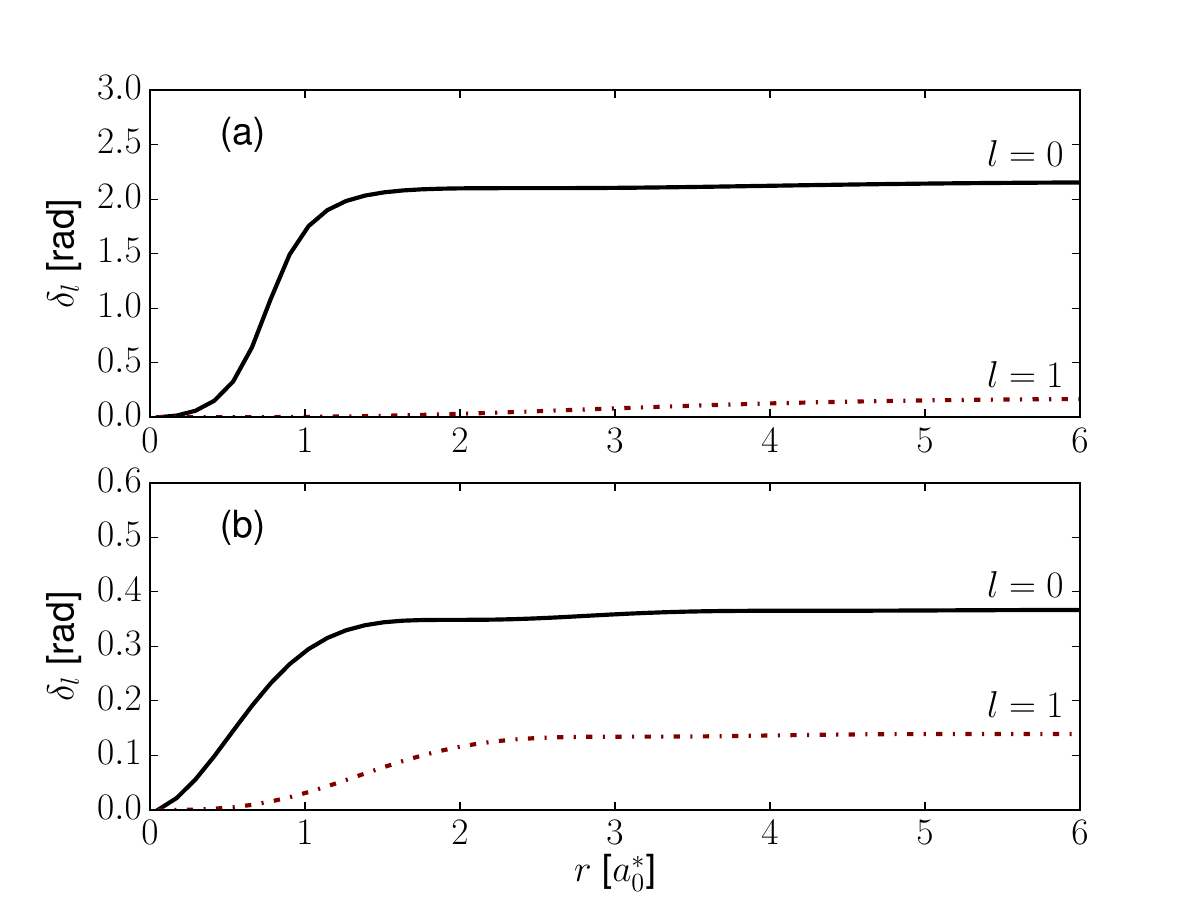}
}
\caption{(a) Phase shifts $\delta_l$ curves  ($l=0,1$) versus inter-particle distance $r$ in  units of the effective Bohr radius $a_0^{\ast}$ from an impurity center for  {\it n}-type  Si. The scattering phase shifts (in radians) are obtained at the large distance limit, i.e., from $\delta_l$ curve's asymptotes.  Here $n_i =  10^{17}$ $\mathrm{cm}^{-3}$ and hence $E_{\mathrm{F}}\approx 3 $ $\mathrm{meV}$. (b) The same as the top panel (a) but for   $n_i =  10^{20}$ $\mathrm{cm}^{-3}$, and hence $E_{\mathrm{F}}\approx 305 $ $\mathrm{meV}$.}
\label{fig:figure1}       
\end{figure}

In Fig.~\ref{fig:figure1} it is shown how the $\delta_l$ for $l=0,1$ are computed in the variable phase approach for a doped Si semiconductor with  $n_i= 10^{17}$ $\mathrm{cm}^{-3} $ (top panel (a)) and  $n_i=  10^{20}$ $\mathrm{cm}^{-3} $ (bottom panel (b)). First, we note that the all phase shifts are positive due to the attractive interaction potential $V_{ei}$. Second, the $\delta_l$ correctly decrease with increasing  scattering energy $E_{\mathrm{F}}= \hbar^{2} k_{\mathrm{F}}^{2}/2 m^{\ast} $ where $k_{\mathrm{F}} \sim n^{1/3} $. Thus the phase shifts $\delta_l$ ($l=0,1$) on the bottom panel (b),  whose values are $\delta_0=0.36$ and  $\delta_1=0.14$ in radians,  are smaller  compared to those shown on the top panel (a), where $\delta_0=2.15$ and $\delta_1=0.17$.  Indeed  the carriers scatter off the impurities with higher energy increasing the dopant concentration. We also note that for a given  scattering energy,  $\delta_1 \ll \delta_0$ due to the repulsive effect of the centrifugal potential $l\left(l + 1\right)\hbar^{2}/2 m^{\ast}r^{2}$ which becomes stronger for higher $l$ values. The  variable phase method provides the numerical values of the  $\delta_l$, as shown in Fig.~\ref{fig:figure1}, from  the asymptotes of the $\delta_l\left(r\right)$ curves and at the same time  removes their $mod\left(\pi \right)$ issue thus allowing the computation of scattering phase shifts in a unique and unambiguous way \cite{calogero1967}. 
For the validity of the Born approximation, here we adopted the criterion proposed by Morse and Allis  \cite{morse1933, calogero1963, calogero1967}, which states the B1 is a good approximation insofar the phase shifts  are smaller than $\pi/2$. This is certainly the case for a scattering potential of Yukawa type, however it is worth noting that some  short-range  potentials can cause small true phase shifts $\delta_l$ and $\delta_l^{1}$ as well, and yet the Born approximation fails \cite{peierls1979} \footnote{This statement is equivalent to say that $\delta_l^{1} \gg \delta_l$ or $\delta_l^{1} \ll \delta_l$, despite the fact that both $\delta_l, \delta_l^{1}$ are small.}.

\section{Results for the Exact Single-particle Relaxation Time} \label{sec:results}

Along with Ref.~ \cite{dasSarma1985}, we  define the dimensionless quantity $y= k_{\mathrm{F}}/q _{\mathrm{TF}}$ as we wish to compare the ratio $\tau_s^{1}/\tau_s$ against $y$. To this end,  we note that $\tau_s^{1}/\tau_s = \sigma_{ei, tot}/\sigma_{ei, tot}^{1} $ where the  exact total  cross-section must be evaluated including a finite number of partial waves necessary to allow the convergence of $\tau_s$ curves in the range of doping densities under scrutiny. Then, the electron-impurity total cross-section is given by

\begin{equation}\label{eq:total}
\sigma_{ei,tot} \left( k_{\mathrm{F}}\right) = \frac{4 \pi}{k_{\mathrm{F}}^{2}} \sum_{l=0}^{l_{max}} \left(2l +1 \right) \sin^{2} \delta_l \, ,
\end{equation}
$l_{max}$ being the maximum of set $\left\lbrace 0,1 \cdots l-1, l \right\rbrace$ for a given $l$, which is expected to be a small integer. In  fact, even on the basis of a semi-classical analysis of scattering, it can be shown that only a few partial waves will actually contribute to the total cross-section in low energy collisions \cite{marchetti2017}. Moreover, by means of Eq.~\ref{eq:totalB1}, we are able include  the potential contribution of all the  partial waves, i.e., for  $l=0, ..\infty$, to the total cross-section in B1, a task that would be proved very inaccurate, if instead we determine $\sigma_{ei,tot}^{1}$ by means of Eq.~\ref{eq:total} and  of the phase shifts in B1 (Eq. \ref{eq:delta1}).

\begin{figure}[htp]
\resizebox{0.55\textwidth}{!}{%
  \includegraphics{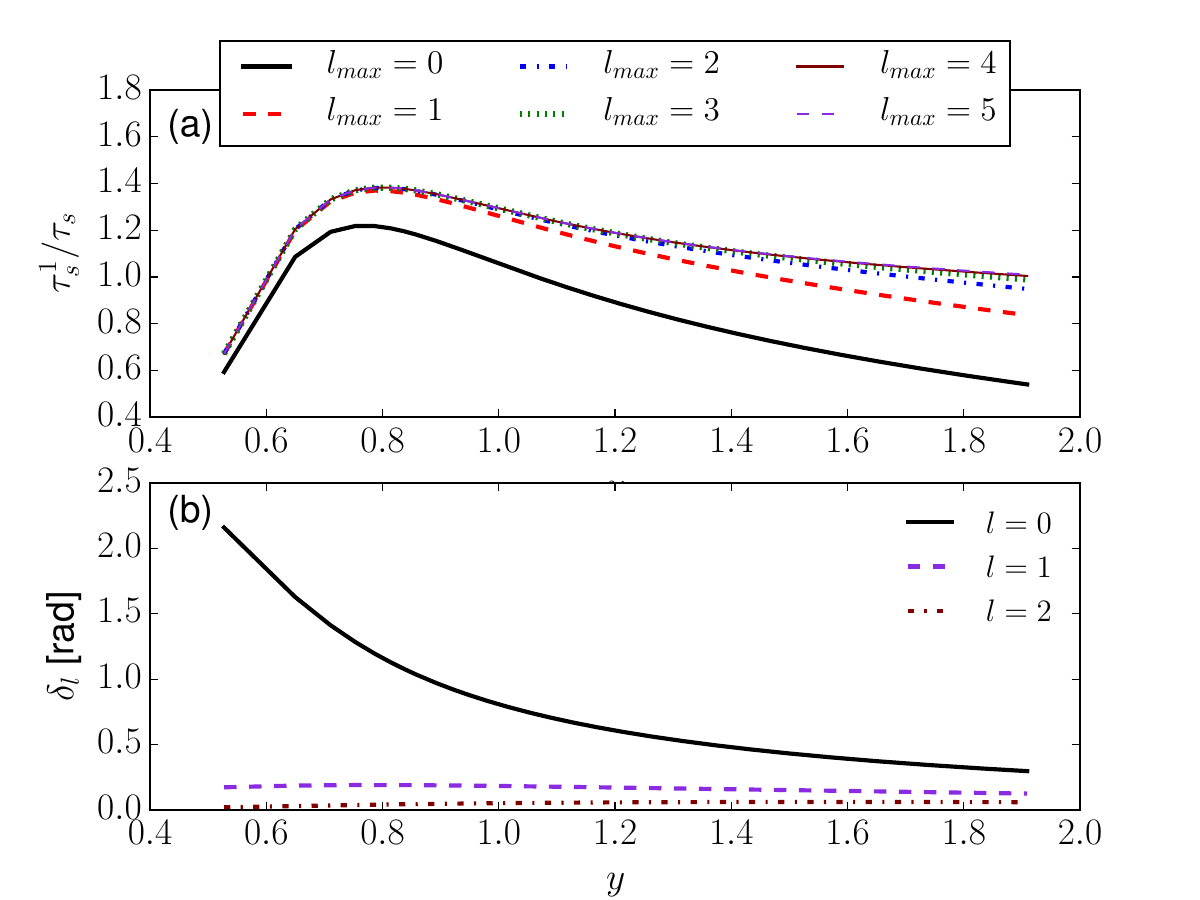}
}
\caption{(a) The curves  $\tau_s^{1}/\tau_s $  versus $y$ for $l_{max} =0, \cdots, 5$ for Si. (b) The relative true phase shifts $\delta_0$,  $\delta_1$ and $\delta_2$ computed by means of the  variable phase approach  versus $y$.}
\label{fig:figure2}       
\end{figure}

\begin{figure}[htp]
\resizebox{0.50\textwidth}{!}{%
  \includegraphics{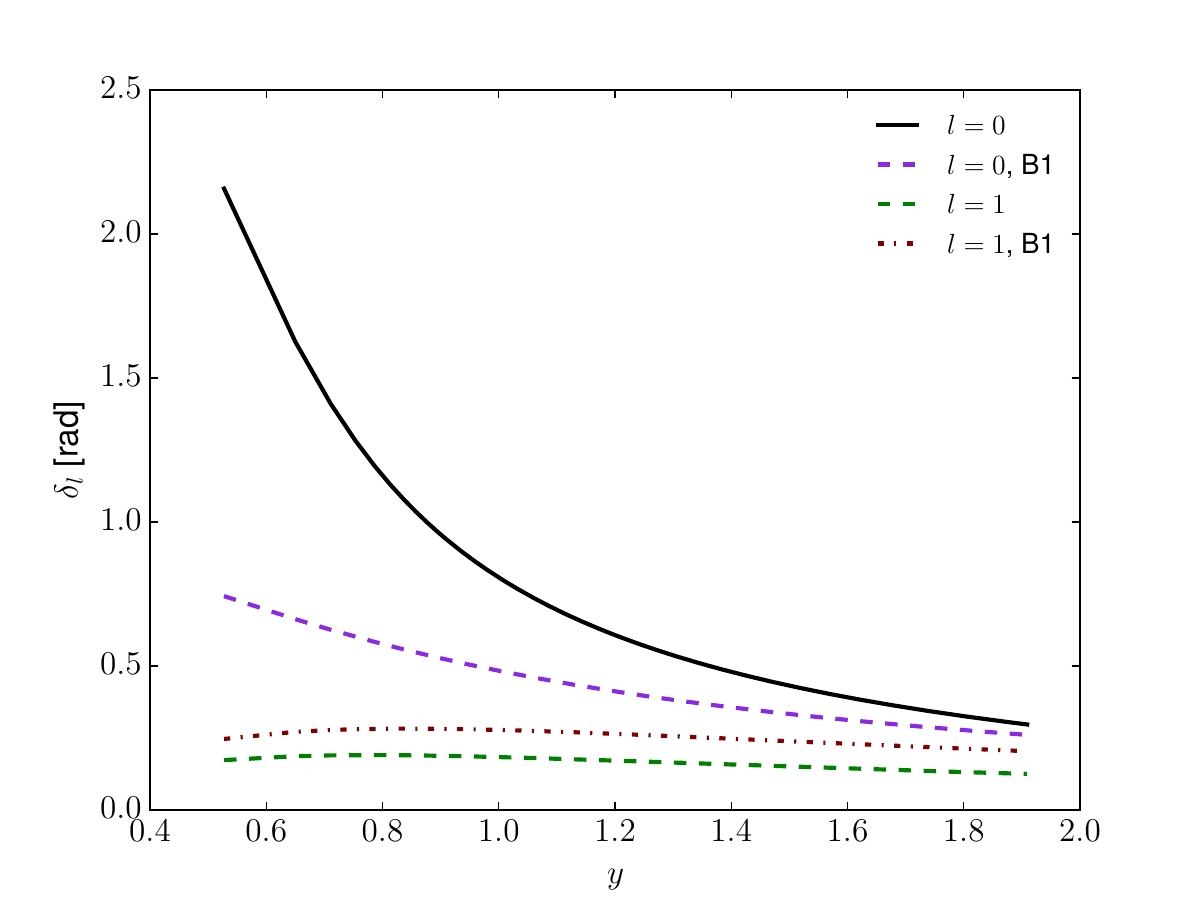}
}
\caption{Comparison between $\delta_l$ and $\delta_l^{1}$, the true phase shifts and those in Born approximation for {\it n}-type  Si with  $n_i= 10^{17} - 2.5 \times 10^{20}$ $\mathrm{cm}^{-3} $  and $l=0,1$.}
\label{fig:Born}       
\end{figure}

In Fig.~ \ref{fig:figure2} (top panel (a)), we plotted the ratio $\tau_s^{1}/\tau_s $ against $y$ for many different $l_{max}$ values for a doped Si semiconductor. Not surprisingly  the major contributions come from $l_{max} =0$ (s-wave, $l=0$) and $l_{max} =1$ which accounts for s-wave and p-wave ($l=1$) together. Their contributions are substantial for $y \ll 1$, the interval which defines the low-energy limit  of scattering theory for short-range potentials. The maxima of the curves happen roughly  at the cross-over region between low-energy and high-energy carrier-impurity collisions, the latter region being defined for $y \gg 1$. The curves clearly show that in B1, the single-particle relaxation can be underestimated by roughly by 30\%   and overestimated  by  roughly 40 \% for $y \lesssim 1$. Such large discrepancies for the single-particle relaxation times  can be understood looking at the relative exact phase shifts computed in the  variable phase approach. In Fig.~ \ref{fig:figure2} (bottom panel (b))   the relative phase shifts $\delta_0$, $\delta_1$, $\delta_2$ are plotted against $y$. Clearly,  $\delta_0$ is too large for the Born approximation to hold: in fact, its validity would require that $\delta_l$ be small compared to $\pi/2$ \cite{bethe1957,calogero1963,morse1933}. However, a better agreement between  $\tau_s^{1}$ and $\tau_s $  is found in the limit $y \to 2$ where  $\tau_s^{1} \approx  \tau_s $. This is consistent with the fact that $\delta_0$, and the other phase shifts as well, decrease monotonically as the carrier energy increases, thus improving the over Born approximation. In this regard, the computation of the discrepancies $\Delta\delta_l =|\delta_l^{1} -\delta_l|$ for $l=0,1$ by means of Eq.~\ref{eq:delta1}, confirms the previous analysis. In Fig.~ \ref{fig:Born},   we compares the true phase shifts and the ones in B1. It it evident that B1 fails for  $y \to 0.4$ as $\Delta\delta_0 \approx 1.41$ (rad) while  $\Delta\delta_l \approx 0$ for $l=0,1$, in the limit of  $y \to 2$ i.e., $\delta_l^{1} \approx \delta_l$, the latter being the condition for validity of the Born approximation.

\begin{figure}[htp]
\resizebox{0.55\textwidth}{!}{%
  \includegraphics{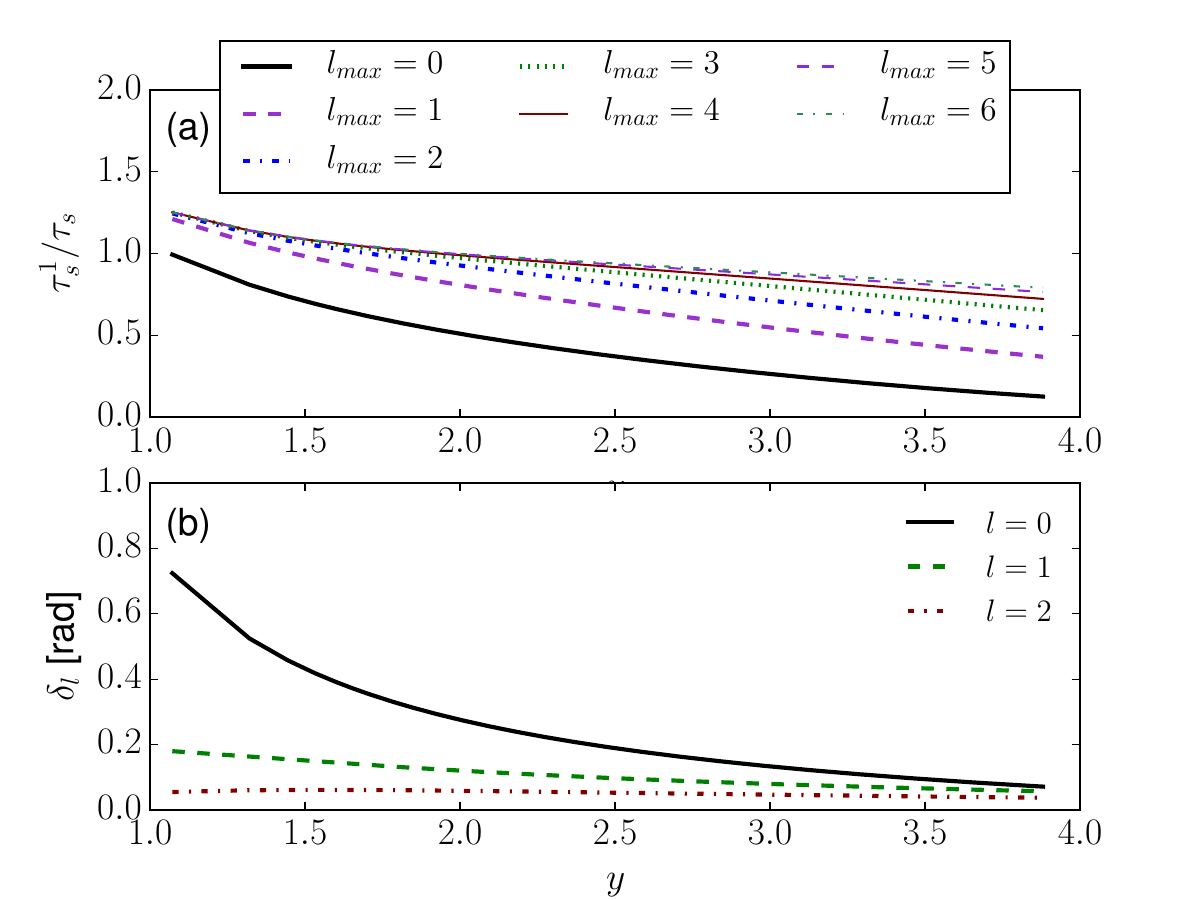}
}
\caption{(a) The curves  $\tau_s^{1}/\tau_s $  versus $y$ for $l_{max} =0, \cdots, 6$ in the case of GaAs. (b) The relative true phase shifts $\delta_0$,  $\delta_1$ and $\delta_2$ computed by means of the  variable phase approach  versus  $y$.}
\label{fig:figure3}       
\end{figure}

In order to understand the semiconductor many-body effects on $\tau_s$, we  performed similar computations for  {\it n}-type  GaAs semiconductor in  the same range of doping concentrations taking the following material parameters: $m^{\ast} =0.067m_e$ and $\kappa=12.9\varepsilon_{0}$ \cite{marchetti2017}. Note that even in GaAs for the dopant concentrations under scrutiny, it is usually assumed that only carrier-single impurity ion collision can occur \cite{lowney1991}.

In  Fig.~ \ref{fig:figure3} (top panel (a)), we show our results for  $\tau_s^{1}/\tau_s $  computed via the VPM. The ratio curves now show a good agreement for a large interval of $y$ values, see top panel (a). We found that $\tau_s^{1}$   can be overestimated  by  roughly 30\% for $y \approx 1$  and underestimated by roughly by 20\% in the limit $y \to 4$. For $ 1.8 \lesssim y \lesssim 2.2 $ the ratio discrepancy is  less than about 3\%. However, here we need to make an important observation about our results  for large $y$ values. It can been shown that for $y \gg 1$ the Born approximation would produce the same results as a bare Coulomb potential (Rutherford scattering) for scattering angles $\theta  \gg 1/y $, see Ref.\cite{haar2014} for an extensive analysis,  thus disregarding the screening effects altogether. Therefore in this case, it is not clear whether or not the Born approximation can be still considered suitable for modeling screened electron-impurity collisions. In particular, this would certainly affect the physics of GaAs for most of the dopant concentrations under scrutiny.  

Nevertheless, the  nicer behavior of  $\tau_s^{1}$  observed in GaAs can be clearly understood from the bottom panel (b) of Fig.~\ref{fig:figure3}, which shows that all phase shifts are now much smaller than  $\pi/2$, thus improving the accuracy of the Born approximation. Note that carriers in GaAs have much more energy available in the center of mass, hence making $\delta_l$ smaller in comparison to those computed for Si. This is a direct consequence of a much smaller effective mass of the carriers in GaAs.

\begin{figure}[htp]
\resizebox{0.40\textwidth}{!}{%
  \includegraphics{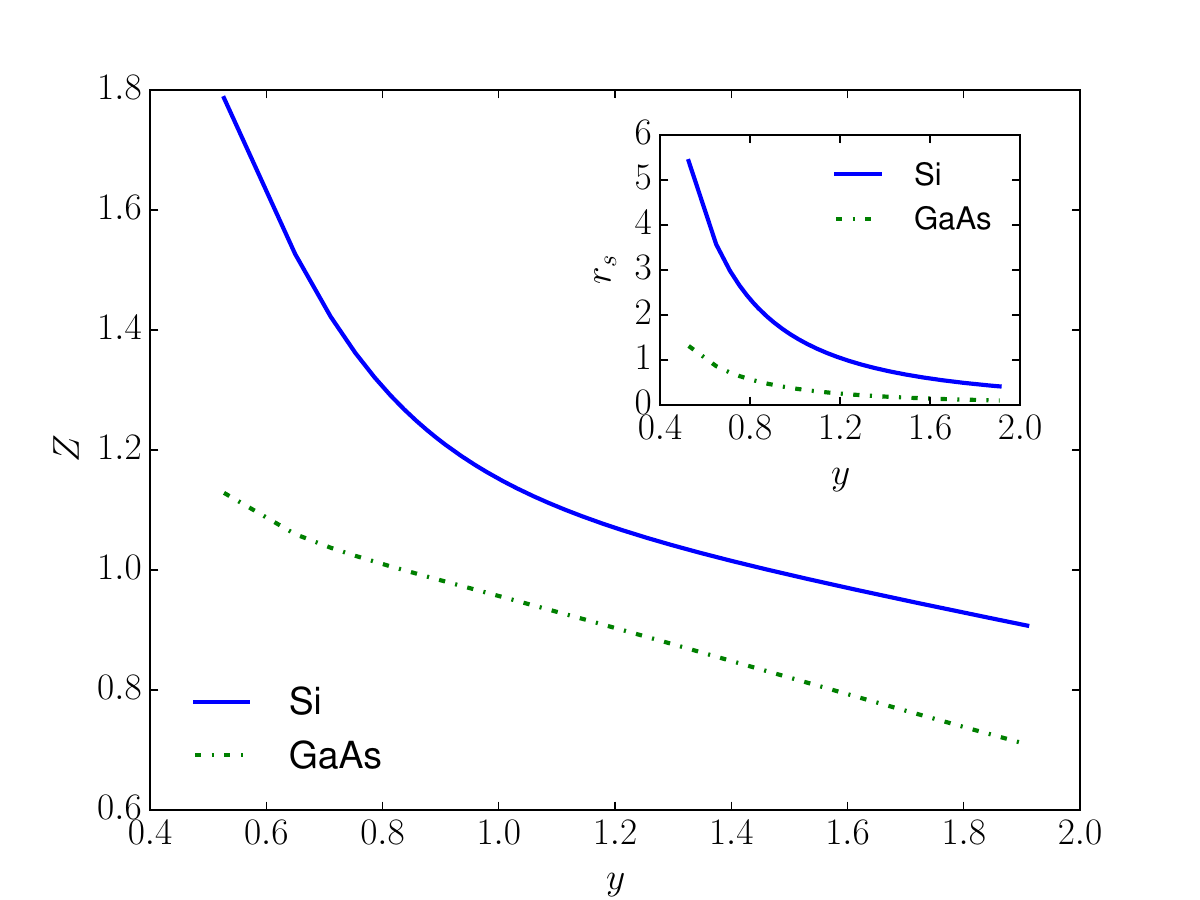}
}
\caption{ The curves  (solid line, Si and dashed line, GaAs) versus $y$  obtained by computing $\delta_l$ from VPM and using Eq.~\ref{eq:fsr} and assuming  $Z=1$. In the inset the relative curves of the dimensionless Wigner-Seitz parameter $r_s$ versus  $y$   are shown.}
\label{fig:figure4}       
\end{figure}

Until now, our findings rely on the validity of the screened Coulomb potential for modelling the electron-impurity interaction in the bulk of doped semiconductors.  Within the present formalism, we can address its relationship with the random phase approximation linking the $\delta_l$ to the Fermi sphere through the  Friedel sum rule (FSR) \cite{mahan2000}. The FSR states that the impurity charge must be completely neutralized by the carriers, and at the same time the extra electrons required to this end,  should fill  the levels up to the Fermi energy of an ideal crystal. In mathematical terms the FSR reads as
\begin{equation}\label{eq:fsr}
Z = \frac{2}{ \pi} \sum_{l=0}^{\infty} \left(2l +1 \right) \delta_l \, .
\end{equation}

By the variable phase approach we computed the scattering phase shifts with   the following cut-offs for the quantum number $l$: $l_{max} = 5$ and $l_{max} = 6$ for Si and GaAs respectively. In Fig.~ \ref{fig:figure4}, we plot the  curves (solid line, Si) and (dashed line, GaAs) relative using Eq.~\ref{eq:fsr} against $y$ in the range $ 0.4 < y < 2$. They monotonically decrease with $y$, as the Fermi energy increases with it as well.  It is  also evident that for Si a strong violation of the Friedel sum rule occurs for $ y \lesssim 0.9 $. This  is indeed a region of low-density electron gas, suggesting that there may be some problems relative to random phase approximation. Whether or not RPA is applicable,  depends on the smallness of the Wigner-Seitz parameter $r_s$ \cite{giuliani2005}. It is expected that  RPA works well for $r_s < 1$ \cite{giuliani2005}.
In the inset of Fig.~ \ref{fig:figure4},  the $r_s$ curves (solid line, Si) and (dashed line, GaAs) in the same range  of $y$ values are shown. Remarkably, the strong violation of Friedel sum rule observed in Si can be linked to the non-applicability of RPA ($r_s \gg 1$) for roughly the same $y$ values. Hence, in that region, it may be necessary to account for the short-range  exchange and correlation effects in carrier dynamics in Si, which are not present in the RPA. Furthermore we observe that for large  $y  \to 2 $, see Fig.~ \ref{fig:figure4}, some large violations of the FSR for Si and GaAs as well would start to happen again. Indeed, according to the present physical model,  there is no way to prevent $\delta_l$ from decreasing for increasing $y$ values. Clearly, these strong violations of the Friedel sum rule suggest that the screened Coulomb interaction of Yukawa form, is no longer reliable. In this regard, Mahan in his many-body analysis of band-gap narrowing in Si at zero temperature, found that assuming the electron-impurity potential of Yukawa form leads to very inaccurate energy terms \cite{mahan1980}.
On the other hand, due to the  smallness of $r_s$, if we wish to continue working in the RPA, and thus keeping the  screened Coulomb potential of this form,  we then would need to include a refinement to the Thomas-Fermi screening parameter  $q_{\mathrm{TF}}$, which requires a self-consistent calculation of it associated to the Yukawa potential ensuring that the Friedel sum rule is satisfied \cite{chattopadhyay1981, lowney1991,bennett1992}.
This further correction is well beyond the scope of this paper, and it is related to the difficult problem of modelling the effective interaction of a test charge embedded in an electron liquid within the RPA \cite{giuliani2005, giuliani2005a}. Note that even within the random phase approximation, a  different screened electron-impurity potential emerges, when the many-body effects are taken into account by the Lindhard screening. In this case, the scattering potential, whose analytical form is not given, shows a tail with an oscillatory behavior far away from the impurity center \cite{ashcroft1976, giuliani2005a}. 

In the present work we limited ourselves to a linearized Thomas-Fermi screening theory, thus ignoring that it is indeed a crude approximation \cite{marchetti2018}. However, despite this fact, it is still widely employed in computations of  material properties \cite{caruso2016}.

\section{Interference of Partial Waves in the Strong Screening Limit} \label{sec:results1}

In the following we show that the strong  interference of partial waves plays an important role at low energies.  This physical effect  is completely missed  in the Born approximation for the problem at hand, and  and it is a direct consequence of a poor scattering approximation when the  scattering phase shifts become large.

In the paper ~\cite{dasSarma1985} Das Sarma and Stern found that in a three-dimensional impure electron gas, when the Thomas-Fermi screening becomes strong, i.e. when $y \ll 1$, the ratio $\tau_t/\tau_s 	\sim 1 $. The reason is that the electron-impurity scattering becomes nearly isotropic for small $y$ values. This is easily understood if one recalls the formula of the differential cross-section $\sigma_{ei}^{1}\left( \theta \right)$, which reads as 

\begin{equation}\label{eq:angular1}
\begin{split}
\sigma_{ei}^{1}\left(\theta\right)& = \frac{4 m^{\ast^2} U_0^{2}}{\hbar^{4} q _{\mathrm{TF}}^{4} \left(4y^{2}\sin^{2}\left(\theta/2 \right)  + 1\right)}   \, .
\end{split}
\end{equation}

Das Sarma and Stern's result is obtained in the Born approximation which  is implicitly assumed in their paper \cite{dasSarma1985}, however the B1 does not take into account the importance of partial wave interference at low energy. To prove this, we computed $\tau_t/\tau_s $ by means of VPM in the doping region $n_i= 10^{16} - 10^{18}$ $\mathrm{cm}^{-3} $ corresponding to $y$ values in the interval $(0.4, 1)$ where  our previous analysis showed that the Born approximation is invalid \footnote{We refrain from including lower dopant concentrations in the computations as the semiconductor physics can become very different from that of our model.}. Note that for computing  $\tau_t $  in term of phase shifts we use the formula \cite{bennett1992, chattopadhyay1981}

\begin{equation}\label{eq:taut}
\tau_t = \frac{4 \pi \hbar n_i}{ m^{\ast} k_\mathrm{F}} \sum_{l=1}^{\infty} l \sin^{2}\left(\delta_{l-1} - \delta_l \right)  \, .
\end{equation}

\begin{figure}[htp]
\resizebox{0.50\textwidth}{!}{%
  \includegraphics{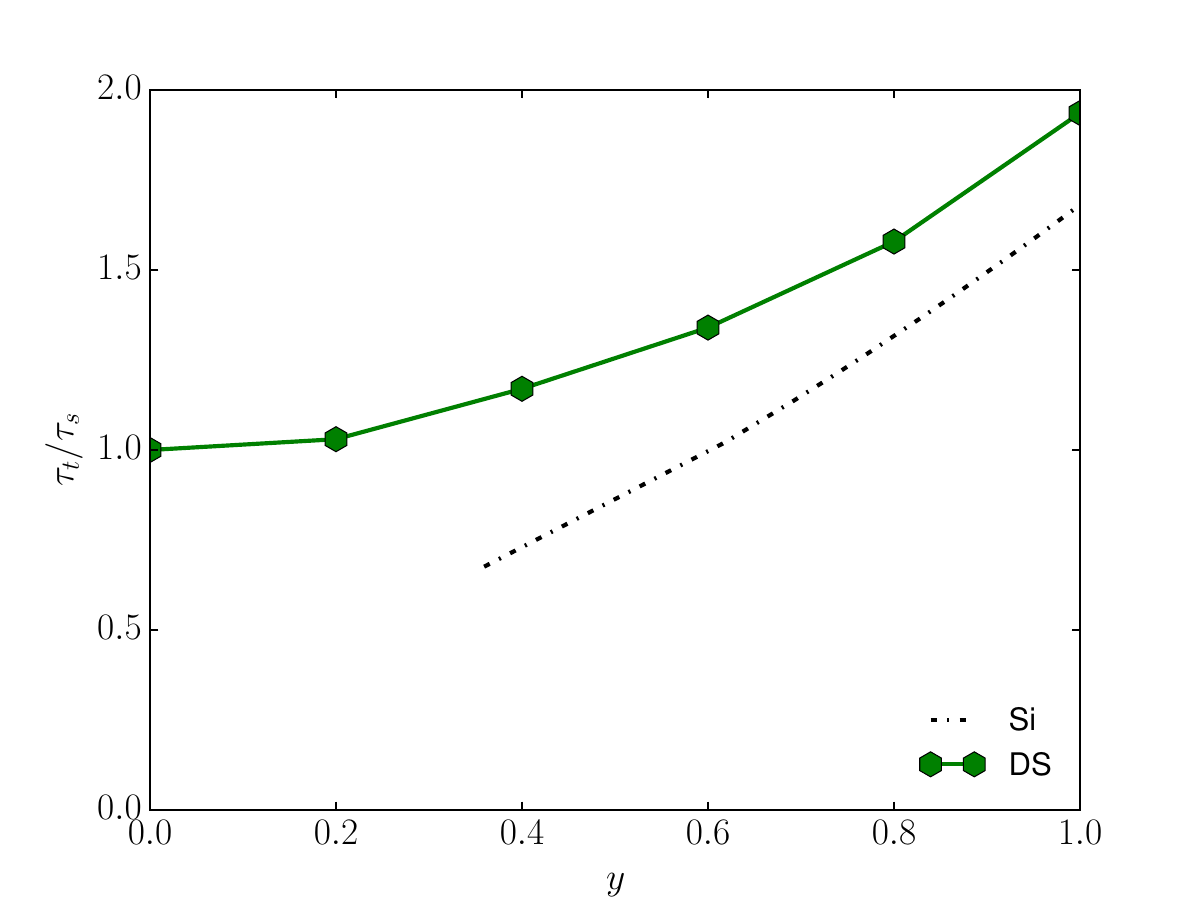}
}
\caption{ The dashed line curve represents  the  $\tau_t/\tau_s$ values versus  $y$ computed by VPM for Si for the doping density interval  $n_i= 10^{16} - 10^{18}$ $\mathrm{cm}^{-3} $, while the solid line curve is the same but obtained in the Born approximation for a three-dimensional semiconductor in the Thomas-Fermi screening approximation, see Ref.\cite{dasSarma1985}.}
\label{fig:dasSarma}       
\end{figure}

In Fig.~ \ref{fig:dasSarma} we compare our results (dashed line curve) for $\tau_t/\tau_s $ versus $y$ with the expected ones (solid line curve) from  Ref.~\cite{dasSarma1985}. It is evident
that $\tau_t/\tau_s $ becomes smaller than unity for $y \ll 1$. The reason of this discrepancy, can be understood expanding $\sigma_{ei}\left( \theta \right)$ in terms of Legendre polynomials \cite{schiff1968}. Since $\delta_0, \delta_1$ are appreciably different from zero for Si, this expansion can be truncated retaining three terms only as \cite{schiff1968}
\begin{equation}\label{eq:angular}
\begin{split}
\sigma_{ei}\left(\theta\right)& =\frac{1}{ k_\mathrm{F}^2}\big[ \sin^{2} \delta_0 + 6 \cos\left(\delta_0 - \delta_1\right)\sin \delta_0 \sin\delta_1 \cos\theta\\
 & \quad +9\sin^{2} \delta_1 \cos^{2}\theta  \big]   \, .
\end{split}
\end{equation}

For the parameters under scrutiny,  $\delta_0 > 2$,  $\delta_1 \approx 0.1$ consistent are with the failure of the Born approximation. Therefore a simple numerical check  shows that the coefficients of  $ \cos\theta$, and $\cos^{2}\theta$  in Eq.~\ref{eq:angular}   are not negligible, making the scattering  strongly $\theta$-dependent at  low energies.

In summary, we showed that the VPM outperforms  the Born approximation when it comes to compute $\tau_s$. From a practical point of view,  these  accurate numerical  values may be employed as input to other condensed matter models and/or to applications of the density functional theory (DFT) \cite{hohenberg1964, kohn1965, capelle2006}. From a theoretical point of view, we gained some important physical insights of the many-body dynamics within the homogeneous electron gas model. In particular, we recover the important role of the interference at low energies. Finally, our approach restores the unitarity (probability conservation) which  is manifestly violated by the Born approximation as it fails to satisfy the optical theorem.

\begin{acknowledgments}
We are grateful to Fabio Caruso and  Feliciano Giustino for some useful comments. We are indebted to  Kristjan Kannike, Marco Patriarca and Sean Fraser for reading the manuscript. This work was supported by institutional research funding IUT (IUT39-1) of the Estonian Ministry of Education and Research, by the Estonian Research Council grant PUT (PUT1356).
\end{acknowledgments}

\bibliography{references}{}

\bibliographystyle{apsrev4-1}

\end{document}